\documentstyle[12pt]{article}
\begin{document}
\vspace{-0.2cm}
\begin{flushright}gr-qc/9904012
\end{flushright}
\vskip1cm
\begin{center}
{\bf KERR--SCHILD  APPROACH  TO THE\\
\vskip3mm
BOOSTED KERR SOLUTION}\\
\vskip1cm
{\bf Alexander Burinskii} \footnote{ e--mail: grg@ibrae.ac.ru}
\vskip2mm
{\em Gravity Research Group, NSI, Russian Academy of Sciences,\\
B.Tulskaya 52, 113191 Moscow, Russia}
\vskip.2in
{\bf Giulio Magli}  \footnote{ e-mail: magli@mate.polimi.it}
\vskip2mm
{\em Dipartimento di Matematica del Politecnico di Milano,\\
Piazza Leonardo Da Vinci 32, 20133 Milano, Italy}
\end{center}
\vskip1cm
\centerline{\bf Abstract}
\par
\begin{quotation}
Using a complex representation of the Debney--Kerr--Schild (DKS)
solutions and the Kerr theorem we analyze the boosted Kerr geometries and
give the exact and explicit expressions for the metrics,
the principal null congruences, the coordinate systems and the location of
the singularities for arbitrary value and orientation of the boost with
respect to the angular momentum.
\par
In the limiting, ultrarelativistic case we obtain light-like
solutions possessing diverging and twisting principal null
congruences and having, contrary to the known pp-wave limiting solutions,
a non-zero value of the total angular momentum.
\par
The implications of the above results in various related fields are discussed.
\end{quotation}
\newpage
\def\b{\bar}
\def\d{\partial}
\def\D{\Delta}
\def\cD{\Cal{D}}
\def\f{\varphi}
\def\g{\gamma}
\def\G{\Gamma}
\def\l{\lambda}
\def\L{\Lambda}
\def\M{{\Cal M}}
\def\m{\mu}
\def\n{\nu}
\def\p{\psi}
\def\q{\b q}
\def\r{\rho}
\def\t{\tau}
\def\x{\phi}
\def\X{\~\xi}
\def\~{\tilde}
\def\h{\eta}
\def\z{\zeta}
\def\Z{{\b\zeta}}
\def\Y{{\b Y}}
\def\cZ{{\b Z}}
\def\`{\dot}
\par\noindent
{\bf 1. Introduction}
\par
\bigskip
Recently, boosted black hole solutions attracted a renewed
interest in connection
with numerical simulations of black hole interactions \cite{[1]}.
On the other hand, the problem of finding
the ultrarelativistic limit of exact, particle-like
solutions of the Einstein field equations
received considerable attention in connection with non-trivial
gravitational effects which are expected to occur in the interparticle
interactions at extreme energies due to the presence of gravitational
shock waves \cite{[2],[3]}.
\par
First results in this field were obtained by Aichelburg and
Sexl \cite{[4]}, who considered the behaviour of the Schwarzschild
metric under ultrarelativistic boost.
Because of spherical symmetry the
results are in this case independent on the direction of the boost.
\par
A similar treatment in the case of the Kerr geometry, which can be considered
as a model of a spinning particle in General Relativity, has to take
into account the orientation of the angular momentum with
respect to the boost.
    Unlike the simplest twist free case the problem exhibits extra
difficulties which enforces quite complicated and refined methods of
analysis and usually leads to very complicated or approximate expressions
before taking the ultrarelativistic limit \cite{[5],[6],[7]}.
In particular, difficulties appear when ultrarelativistic limits are
involved, due to the singular character of Lorentz transformations at $v=c$.
We should note that this singularity a priori can lead to different
limiting results depending on the performed limiting procedure.
\par
The approach which we are going to formulate here  deals with a
class of {\it sourceless} gravitational solutions of Kerr-Schild class
and it is based on the Debney, Kerr and Schild formalism \cite{[8]} (DKS)
and on the Kerr theorem \cite{[9],[10],[11]}.
It gives the possibility of obtaining exact and
explicit expressions for the boosted Kerr geometry by arbitrary values
and orientations of the boost with respect to the angular momentum.
As a result, in the general cases of the boost, including the
ultrarelativistic cases, we determine the  exact expressions for the
metric, coordinate system, principal null congruence, and
location of singularity.
In the ultrarelativistic cases this method leads to a DKS-class of
solutions possessing diverging and twisting principal null congruences
contrary to the pp-wave limiting solutions which posses zero total
angular momentum \cite{[7]}.
\par
The paper is organized as follows.
First, we briefly recall the
DKS- formalism and the Kerr theorem in a form, in which
the Kerr solution is represented as being generated by a complex source.
This approach was initiated by Lind and Newman \cite{[12],[13]} and was
considered in the DKS-formalism in [14] (a more complete description of
this approach and of its geometrical basis can be found in \cite{[15]}).
Using this representation,
the boosted Kerr solutions can be constructed simply
by considering straight lines
in complexified Minkowski space
as complex world lines of the sources.
In this way,
we obtain explicit expressions for the metric and the
singular regions in the most representative
cases. We discuss various physical applications and some unusual features
in the limiting behaviour of the Kerr singular ring.
\bigskip
\par\noindent
{\bf 2.  The DKS formalism and the Kerr theorem}
\par
\bigskip
In the notation we follow the work of Debney, Kerr and Schild \cite{[8]}
(see also \cite{[9],[16]} for a review on DKS solutions).
In the four-dimensional space-time with signature  $(-+++)$, let
$ e_1, e_2, e_3, e_4 $ be a null tetrad satisfying
\begin{equation}
g_{ab}=
e_a^\m e_{b\m}=
\left( \begin{array}{cccc}
0&1&0&0 \\
1&0&0&0 \\
0&0&&1 \\
0&0&1&0
\end{array} \right)
\end{equation}
The vectors $e^3$ and $e^4$ are real null
vectors, while $ e^1$ and $e^2 $ are complex conjugates.
The general Kerr-Schild metric can be written as
\begin{equation}
g_{\m\n} =
\h_{\m\n} + 2 h e^3_{\m} e^3_{\n} \ ,
\end{equation}
where $h$ is a scalar function and
the principal null direction $ e^3 $
is null also with respect to
the auxiliary Minkowski space with metric
$$
\h = dx^2 + dy^2 + dz^2 -dt^2 = 2du dv+2d\Z d\z \ .
$$
In the above formula,
the null coordinates $\z ,\Z , u ,v$
are related to the Cartesian coordinates by
\begin{eqnarray}
2^{1\over2}\z &=& x+iy ,\qquad 2^{1\over2} \Z = x-iy ,\nonumber \\
 2^{1\over2}u &=& z + t ,\qquad 2^{1\over2}v = z - t  .
\end{eqnarray}
A general field of null directions in Minkowski space
can be defined by
\begin{equation}
e^3 = du+ \Y d \z  + Y d \Z - Y \Y d v \ ,
\end{equation}
where $Y(x)$ is a complex function.
The Kerr theorem gives a rule to construct the geodesic,
shear free congruences:
the general geodesic, shear-free null congruence in
Minkowski space is defined by a function $Y$ which is a solution of the
equation
\begin{equation}
F (\l_1,\l_2 ,Y) = 0 \ ,
\end{equation}
where  $F$ is an arbitrary analytic
function of the {\it projective twistor coordinates}
\begin{equation}
\l_1 = \z - Y v, \qquad \l_2 =u + Y \Z, \qquad Y\ ,
\end{equation}
(the above parameters can be written
in twistor notation  $(\m ^A , \p _{\dot A}),$
$\m ^A = x^\m \sigma _\m^{A \dot A} \p _{\dot A}$ as
$ ( \l_1, \l_2, Y, 1) =
(\m^0,\m^1,\p_{\dot 0},\p_{\dot 1})/\p_{\dot 1}.$)
\par
The Kerr theorem also allows to obtain other important parameters of
the solution.
In particular,
the quantity
\begin{equation}
\tilde r := - d F / d Y \ ,
\end{equation}
is a complex radial distance,
which is connected with the complex representation of
the Kerr solution, as will be explained below.
\par
The
singularities of the metric  can be defined as the
caustics of the congruence given by the system of equations
\begin{equation}
F=0, \qquad
d F / d Y =0 \ .
\end{equation}
The Kerr solution belongs to the sub--class of
solutions having singularities
contained in a bounded region \cite{[14],[17]}.
In this case the function $F$ must be at most quadratic in $Y$:
\begin{equation}
F \equiv a_0 +a_1 Y + a_2 Y^2 + (q Y + c) \l_1 - (p Y + \q) \l_2 \ ,
\end{equation}
where the coefficients $ c$ and $ p$ are real constants
and $a_0, a_1, a_2,  q, \q, $  are complex constants.
The solutions of the equations (8)
can be found in this case in explicit form. It can be shown that
they correspond to the Kerr solution up to a Lorentz boost and a
shift of the origin.
\bigskip
\par\noindent
{\bf 3. Congruences generated by a complex world line}
\bigskip
\par\noindent
Our approach is based on the complex world-line representation of the
Kerr solution which was initiated by Newman and Lind \cite{[12],[13]}.
It allows to represent the Kerr solution as a retarded-time field,
generated by a complex source propagating along a complex world-line.
The structure of this representation in DKS-formalism
gives a very convenient form of the function $F$,
which turns out to be dependent on
the coordinates of the world-line \cite{[14],[15]}.
For the aim of convenience we give here a short review of this approach.
\par
Let $ x_0^\m (\t)$
be a complex world line
parameterized by a complex time parameter $\t=t+i\sigma$. The
coordinates of this world line are complex,
$ x_0 (\t)= (\z_0, \Z_0, u_0 ,v_0) \in CM^4,$ so that
$\Z_0$ and $\z_0$ are not necessarily complex conjugates.
\par
The function $F$ can be expressed in the form
\begin{equation}
F \equiv (\l_1 - \l_1^0) \widehat K \l_2 - (\l_2 -\l_2^0) \widehat K \l_1.
\end{equation}
where the twistor components with zero indices
\begin{equation}
\l_1^0 (\t)= \z_0(\t) - Y v_0(\t), \qquad
 \l_2^0(\t) =u_0(\t) + Y \Z_0(\t),
\end{equation}
denote the values of $\l_1$ and $\l_2$
on the points of the complex world-line
$ x_0 (\t)$, while
$\widehat K$ is a Killing vector of the solution,
whose action on a scalar $f$ is
defined by
\begin{equation}
\widehat K f= \`x_0^\m(\t) \d_\m f,
\end{equation}
(a dot denotes derivative with respect to $\tau $).
\par
It has been shown [15] that the form of $F$ given by (10)
is equivalent to (9).
In this representation, the Kerr congruence
can be described via a retarded-time
construction.
For example,
the Schwarzschild and the Kerr metrics correspond to
the world line of a particle at rest at the origin and
to the world line of a particle at rest ``at a distance
$ia$ from the origin'', respectively \cite{[13]}.
The general case
of a solution with a boost may be
obtained considering a straight complex
world line with 3-velocity $\vec V$ in $CM^4$
\begin{equation}
x_0^\m (\t) = x_0^\m (0) + \xi^\m \t; \qquad \xi^\m = (1,\vec V)\ .
\end{equation}
Writing the function F in the form
\begin{equation}
F = A Y^2 + B Y + C,
\end{equation}
where
\begin{eqnarray}
 A &=& (\Z  - \Z_0) \`v_0 - (v-v_0) \`\Z_0 ;\nonumber\\
 B &=& (u-u_0) \`v_0 + (\z - \z_0 )\`\Z_0
  - (\Z - \Z_0) \`\z_0 - (v - v_0) \`u_0 ;\nonumber\\
C &=& (\z - \z_0 ) \`u_0 - (u -u_0) \`\z_0,
\end{eqnarray}
from $F=0$ we obtain the following
explicit solutions for the function $Y(x)$:
\begin{equation}
Y_{1,2} = (- B \pm \D )/2A,
\end{equation}
where $\D = (B^2 - 4AC)^{1/2}$.
On the other hand from equations (7) and (14) one
obtains
\begin{equation}
Y = - (B + \tilde r)/2A,
\end{equation}
and consequently
\begin{equation}
\tilde r = \mp (B^2 - 4AC)^{1/2}.
\end{equation}
This relation reflects the ``twofoldedness''
of the Kerr geometry:
the complex radial coordinate
$\tilde r$ can be expressed as
$r + ia \cos \theta $ and the double sign corresponds to a
transition from the "positive" $r$ sheet of the metric to the
"negative" one where $ r \leq 0 $.
\par
In the DKS notation the metric can be written as
\begin{equation}
g_{\m\n} =
\h_{\m\n} + (m/P^3)(Z +\bar Z) e^3_{\m} e^3_{\n}\ ,
\end{equation}
where
\begin{equation}
P =  \`x_o^\m(\t) e^3_\m \ .
\end{equation}
The field $e^3$ can be normalized by introducing $l^\mu = e^{3 \mu}/P$ so
that ${\dot x}_0^\m l_\m =1, $ and this yields the following,
equivalent form of the metric:
\begin{equation}
g_{\m\n} =\h_{\m\n} + m(P^{-1} Z + P^{-1} {\bar Z})
l_{\m} l_{\n}\ .
\end{equation}
The complex radial distance (7) is given by
\begin{equation}
PZ^{-1}=\tilde r \ ,
\end{equation}
and the twistor parameters $\l_1$ and $\l_2 $ may
be represented in the form
\begin{equation}
\l_1 = x^\m e^1_\m , \qquad \l_2 = x^\m (e^3_\m - \Y e^1_\m),
\end{equation}
(the explicit form of the DKS-tetrad $e^a$ is given in the Appendix).
\medskip
\par\noindent
{\bf 4. Boosted Kerr solution: examples and behaviour of singularities}
\medskip
\par\noindent
As we have seen, there is a one-to-one correspondence between
straight lines in complex Minkowski space and
the class of the DKS solutions having singularities contained in a
bounded region.
Since such solutions are equivalent to the Kerr one up
to Poincar\'e' transformations,
it is clear that ``boosting'' the Kerr solution
by a velocity $\vec V$
is equivalent to consider a DKS solution generated
by a particle moving with a speed $\vec V$
in complex Minkowski space.
This motion can be represented in the form
$x_0^\mu (\tau) = \{ \tau, \vec x_0 (0)+ \vec V \t \}$,
and the complex parameter $\t $ may be always chosen
in such a way that $Re \tau$ corresponds to
 the ``real time''$t$ (see \cite{[13]}
for details on complex Minkowski space).
The complex initial displacement can be decomposed as
$\vec x_0 (0) = \vec c + i\vec d$, where $\vec c $ and $\vec d$
are real 3-vectors with respect to
the space O(3)-rotation. The real part $\vec c$
defines the initial shift of the solution, and the imaginary part $\vec d$
defines the size and the position of the
singular ring as well as the corresponding angular
momentum.  It can be easily shown that in the rest frame, when $\vec V=0,
\quad \vec d =\vec d_0 $, the singular ring lies
in the plane orthogonal to
$\vec d$ and has a radius $a=\vert \vec d_0 \vert $. The corresponding
angular momentum is $\vec J = m \vec d_0.$
\par
In the case of a boost
orthogonal to the direction of
$\vec d$, this vector is not altered by Lorentz contraction
($\vec d=\vec d_0$, $\vert \vec d \vert =a$),
while if $\vec d$ and $\vec V$ are collinear we have
\begin{equation}
\vec d_0=\vec d/\sqrt{1-\vert\vec V \vert^2}\ .
\end{equation}
This shows that the parameter $a$ coincides with its rest value $a_0$ if
$\vec d $ and $\vec V$ are orthogonal, while
\begin{equation}
a_0=a/\sqrt{1-\vert\vec V \vert^2}\ ,
\end{equation}
if $\vec V$ and $\vec d$ are collinear.
\par
In order to calculate the parameters $A,B,C$ it is convenient
to express the complex world line in null coordinates
\begin{eqnarray}
2^{1\over2}\z _0 & = & x_0+iy_0 ,
\qquad 2^{1\over2} \Z _0 = x_0-iy_0 ,\\ \nonumber
2^{1\over2}u_0 & = & z_0 + t_0 ,\qquad 2^{1\over2}v_0 = z_0 - t_0 \ .
\end{eqnarray}
The Killing vector of the solution will then be
\begin{eqnarray}
\xi^\mu = 2^{-1/2}\{\` u_0-\`v_0, \` \z_0 +\`\Z_0, -i( \` \z_0 +\`\Z_0),
\` u_0+\`v_0 \},
\end{eqnarray}
while the function  $P$ takes the form
\begin{eqnarray}
P= e^3_\mu {\dot x}_0^\mu =
\` u_0 + \Y \` \z_0  + Y  \` \Z_0 - Y \Y  \`v_0 .
\end{eqnarray}
\par
The complex radial coordinate $\tilde r\equiv PZ^{-1}$ is given
by (18).
As for the standard Kerr solution,
one can represent $\tilde r$
as a ``sum'' of the real radial distance $r$ and an
angular coordinate.
Then equation (18) can be used to fix
the relation between the polar coordinates
$r, \theta, \phi$ and the null Cartesian coordinates (26) through the
expressions (15) for the coefficients $A,B,C$.
Due to the formula (7),
the singular regions are defined by the zeros of the
function $\tilde r.$
In what follows,
we present some examples of boosted Kerr solutions
and then discuss the general features exhibited by them.
\par
\medskip
\noindent
{\sl Example I}
\par
\medskip
\noindent
A spinning particle moves with the speed of light in
the positive direction of the $z $- axis, and the 3-vector
${\vec d}= (0,0,a)$ is also directed along the $z$-axis.
The complex world line is therefore given by
$t_0 (\tau ) = \tau $,
$ z_0 (\tau ) =ia + \tau, \quad x_0(\tau )=y_0(\tau )=0.$
In null coordinates this yields
$$
\sqrt{2} u_0= z_0+\tau =ia+2\tau ; \quad \sqrt{2} v_0= z_0-\tau  =ia, \quad
\z_0=\Z_0=0,
$$
so that $\`u_0= \sqrt{2}, \quad \`v_0 =0, \quad
\` \z_0 =\`\Z_0 =0,$ and
\begin{eqnarray*}
u-u_0 & = & (z-ia +t-2\tau )/\sqrt{2}, \\
v-v_0 & = & (z-ia -t )/\sqrt{2}, \\
\z-\z_0 & = & \z, \qquad \Z-\Z_0=\Z \ .
\end{eqnarray*}
Formula (3.3) implies that the coefficients
$A,B,C$ are given by
$$
A=0;\quad B=t-z+ia;\quad C= x+iy\ .
$$
As a result the function $F$ acquires the form
\begin{equation}
F=x+iy - Y (z-ia -t),
\end{equation}
and the solution of the equation $F=0$ is
\begin{equation}
Y= (x+iy)/(z-ia-t) \ ,
\end{equation}
so that
\begin{equation}
\tilde r = - dF/dY = z-ia-t ,
\end{equation}
and therefore the metric has no singularities
(there is no real
solution to the system of equations (8)).
On the other hand, setting $a=0$
we obtain the case of spinless particle,
and a moving singular plane is placed at $z=t$ \cite{[4]}.
Therefore there is no smooth limit as $a\rightarrow 0$.
\par
\medskip
\noindent
{\sl Example II}
\par
\medskip
\noindent
A spinning particle moves with the speed of light in
the positive direction of the $x$- axis, orthogonal to the 3-vector $\vec d$
which defines the direction and the value of the
angular momentum ${\vec J}= m(0,0,a)$,
$a=\vert \vec d \vert$.
We have the complex world line $ t_0 (\tau) =
 x_0 (\tau ) = \tau ,\quad y_0(\tau )=0,\quad
z_0 (\tau ) =ia\ .$
Correspondingly, the world line in null coordinates is
$$
\sqrt{2} u_0=ia+\tau ,\quad\sqrt{2} v_0=ia-\tau ,\quad\sqrt{2} \zeta_0=\tau,
\quad \sqrt{2} \bar \zeta_0=\tau  \ ,
$$
and the velocities are
$\sqrt{2} \dot u_0=1,\quad\sqrt{2} \dot v_0=-1,\quad\sqrt{2} \dot \zeta_0=1,
\quad \sqrt{2} \dot {\bar \zeta_0}=1.$
We have therefore
\begin{eqnarray*}
\sqrt{2}(u- u_0) &=&z+t-ia-\tau ,
\quad\sqrt{2}(v- v_0)=z-ia-t+\tau,\\
\sqrt{2} (\zeta-\zeta_0)&=&x+iy -\tau ,\quad
\sqrt{2} (\bar \zeta-\bar \zeta_0)=x-iy-\tau  \ ,
\end{eqnarray*}
and the coefficients $A,B,C$ take the form
\begin{equation}
A= (-x+iy -z+t+ia)/2;\qquad B=ia+iy -z;\qquad C= (x+iy -z-t+ia)/2\ .
\end{equation}
The function $Y(x)$ takes the form
\begin{equation}
Y= (x-t-z + iy +ia)/(x-t +z-iy -ia).
\end{equation}
The function $\tilde r\equiv PZ^{-1}$ takes the form
\begin{equation}
PZ^{-1} = -d F/dY = x-t\ .
\end{equation}
This solution is, therefore, singular:
there is a moving singular plane placed at  $x=t.$
\medskip
\par\noindent
{\sl Example III}
\medskip
\par\noindent
The absence of a smooth limit in the
first example may be better understood
considering the general case in which
the value of the velocity
is arbitrary as well as its direction
with respect to the angular momentum.
Without loss of generality,
we can consider
the boost performed with a parameter
$\alpha $ in the $z$-direction $(\alpha=v_z/c)$, and a parameter $\beta$ in
the $x$-direction $(\beta=v_x/c)$, while
the angular momentum is defined by  ${\vec d}= (0,0,a)$.
Denoting
$w^2=\alpha^2 +\beta^2$
the following general formula for the coordinate relations
can be obtained:
\begin{eqnarray}
(x-\beta t)\sqrt{1-\alpha^2} + iy \sqrt{1-w^2} &=&
(r +i a\sqrt{1-\beta^2}) e^{i\phi}\sin\theta,\\ \nonumber
z- \alpha t &=& - r \cos\theta/\sqrt{1-\beta^2}.
\end{eqnarray}
The singular region $r=0, \cos\theta=0$ is placed on the plane
$z=\alpha t$ and is described by
\begin{equation}
\qquad  (1-\alpha^2)(x-\beta t)^2 + (1-w^2) y^2 = a^2 (1-\beta^2).
\end{equation}
Let us first consider the cases corresponding to the two examples above.
If $\beta=0$
(boost in the direction of the angular momentum)
the singularity is
a ring of radius $a_0 =a/\sqrt{1-\alpha^2}$
located on the moving plane $z=-\alpha t$.
In this case, the coordinate relations are the following:
\begin{eqnarray}
x+iy &=& (r+ia) e^{i\phi} \sin \theta /\sqrt{1-\alpha^2},\\
 z-\alpha t &= r \cos \theta.
\end{eqnarray}
The radius of the ring
grows as $\alpha $ increases,
and for $\alpha\rightarrow 1$ the singularity
goes at infinity.
At first sight, this result looks strange.
However, it is easy to check that
being to reexpressed via $a_0$, the rest value of $a$, the position of
singular region becomes a ring of constant radius $a_0$.
\par
However, keeping $a=const.\ne 0$ and taking the limit $\alpha =1$ we obtain
that singular region is going to infinity, so there will not be singularity
in finite region, in agreement with results of Example I.
We will discuss this situation later in connection with the problem of
renormalization of parameters.
\par
In the case $\alpha =0$,
the boost is performed with a speed
orthogonal to  the direction of angular momentum.
The singular region
is a moving ring oblate in the $x$
direction with a Lorentz factor $\sqrt{1-\beta^2}$
and located on the plane $z=0$.
In this case the coordinate relations are:
\begin{eqnarray}
(x-\beta t)/\sqrt{1-\beta^2} + iy &=& (r/\sqrt{1-\beta^2} +ia)
e^{i\phi}\sin\theta,\\
z &=& r \cos\theta/\sqrt{1-\beta^2}.
\end{eqnarray}
As in the previous example,
the limit $\beta\rightarrow 1$ is not smooth
since the singular region is a line
parallel to the $y$ axis placed at $x=t$, $z=0$.
\par
In the general case described by equation (37)
the singularity is a moving ring on the $z=\alpha t$ plane,
deformed in the $x$ direction by a
factor $\sqrt{{(1-\beta^2)}/{(1-\alpha^2)}}$ and in the $y$ direction
by a factor $\sqrt{{(1-\beta^2)}/{(1-w^2)}}$.
The ultrarelativistic limit corresponds to $w=1$
and the singular region is a couple
of straight lines parallel to the $y$ axis.
Therefore,
we can conclude that the non--smoothness and the non--commutativeness
of the limiting procedure is a general feature of the boosted Kerr
solutions.
Another peculiarity, which can be seen by the analysis of the above
examples is a non--trivial coordinate dependence of the function $Y$ which
forms the principal null congruence. As a consequence the congruence
itself acquires a non--trivial coordinate dependence and a non-zero
expansion $\theta$ and twist $\omega$.
This property is conserved even in the
ultrarelativistic limit.
For instance in the case of Example I the expansion and twist of the
congruence are defined by
$ Z= \theta +i \omega $ \cite{[8]} and are given by
$Z/P=-(dF/dY)^{-1} = (z-t +ia)/[(z-t)^2 + a^2]$.
One sees that there is no singularity in this case, and
expansion tends to zero only at the $z=t$ plane where the twist takes
the constant value $1/a$.
\bigskip
\par\noindent
{\bf 5. Concluding remarks}
\bigskip
\par\noindent
There are three different physical situations
which should be described
by the boosted Kerr solution.
\par
The first is connected with the original Aichelburg--Sexl
problem, namely the description of
the gravitational field of light--like particles
with or without spin.
For this case the problem of "renormalization" of
the parameters of the solution has been discussed by many authors
\cite{[5],[7]}.
Indeed the light--like particle must have a infinitely
small rest mass in such a way
that the boosted momentum will be finite.
Similarly, "renormalization" of other parameters, such as charge and
angular momentum, has been discussed \cite{[6],[7]}, and there is no
yet an unique agreement concerning this renormalization procedure.
The above considerations on the behaviour of the singular ring
under the boost in the orthogonal direction
suggest, however, that the physical most satisfactory way
to perform the "renormalization" in this case
should be to keep $J=ma=$const.
In fact in this way the projection of the angular momentum
on the direction of the boost
is invariant with respect to the value of the boost.
One can come to this conclusion also by
considering spinning particles in a quantum context,
since the projection of the spin on the boost direction is
the helicity which indeed must be considered as a constant.
In terms of the rest values $a_0$ and $m_0$
we have to put $J=ma=m_0 a_0 =const.$
As far as $m = m_0/\sqrt{1-v^2}$ by the boost, this
yields $a_0 = a/\sqrt{1-v^2}$.
\par
Therefore, for finite values of $a$ the
rest value $a_0$ and the location of the singularity {\it tend to infinity}
in the ultrarelativistic limit, explaining the results of Example III, and
clarifying the absence of the singularity in the ultrarelativistic limit of
Example I.
\par
One should note that, keeping the value $a_0=const.$ during the limit, one
enforces a fixed size of the singular ring, so that putting
in the limit $m_0 =0 $ one obtains in fact a limiting twist-free solution
with $J=m_0 a_0\rightarrow 0$. This corresponds to the known results
on pp-wave limiting metrics\cite{[5],[6],[7],[18],[19]} with a finite
size of the singular ring and, correspondingly, vanishing total angular
momentum \cite{[7]}.
\par
These arguments are not valid for Example II since the corresponding
projection of angular momentum is initially zero, and we have
here a twist-free solution with a finite location of singularity.
\par
The second application,
which has lead to the recently renewed interest in this problem,
consists in modelling the gravitational field
of elementary particles with finite rest mass
under the boost.
A specific feature of this case is that the rest mass $m_0$ as well as
the projection of the angular momentum $J=m_0 a_0=ma$ have to be kept
constant. This leads to $m_0= const.$, and consequently we obtain
a finite position of singularity which is determined by the value of $a_0$.
\footnote{One sees that singular ring is not subjected to Lorentz
contraction in this case since it lies in the plane
orthogonal to the boost direction.}
However, the parameter of the solution $a$ has to be scaled by the boost
as $a = a_0 \sqrt{1-v^2}$.
\par
In both problems described above one deals with naked singularities
rather than black holes, since the values of mass, spin and charge
of elementary particles typically correspond to this
kind of solutions.
\par
The third physical problem is connected with
astrophysical applications \cite{[1]} of the boosted black hole solutions.
In this case also the behavior of the horizon and of the ergosphere
under the boost are of interest.
A simple analysis using the above suggested coordinates
shows that the horizon as well as the ergosphere
are simply given by the known formulae for the
Kerr case where $m$ must be the relativistic mass parameter.
\par
The method proposed here allows to describe in {\it explicit}
form the metric and
the behaviour of the singular region of the Kerr solution under arbitrary
boost and with arbitrary orientations of the angular momentum.
In particular, we have shown that
the Kerr theorem automatically allows to obtain the exact form of the
boosted  solution in an asymptotically flat coordinate system
and the equations describing  the singularities in these coordinates.
The ultrarelativistic limit is a singular point of the Lorentz
transformations, and we have obtained a quite general
picture of the non-smoothness and non-commutativeness of
the limits $a \rightarrow0$, $v \rightarrow 1$ and $r\rightarrow 0$.
The method shows that light-like limits
of the Kerr geometry exist
which belong to DKS-class and have twisting principal
null congruences and non-zero total angular momentum $J$.
The results can be easily extended also to the boost of  the
Kerr-Newman solution and the Kerr-Sen \cite{[20]} solution generalizing
the Kerr solution to low energy string theory. It was shown in \cite{[21]},
the one of the principal null congruences retains its properties to be
geodesic and shear free , and that the Kerr theorem remains valid for the
Kerr-Sen solution too \cite{[22]}.
\bigskip
\par\noindent
{\bf Acknowledgements}
\bigskip
\par\noindent
The authors gratefully acknowledge Prof. Elisa Brinis Udeschini for
interesting discussions.
One of us (A.B) is grateful to Prof. Elisa Brinis Udeschini for hospitality
at Politecnico di Milano and to G. Alekseev for useful conversations.
\bigskip
\par\noindent
{\bf Appendix}
\bigskip
\par\noindent
Let $e^a$ be a null tetrad and define
the Ricci rotation
coefficients as
$$
\G ^a_{bc} = - e^a_{\m;\n} e_b^\m e_c^\n.
$$
The principal null congruence
has the  $e^3$  direction as tangent.
It will be geodesic if and only if $\G_{424} = 0$ and shear
free if and only if $\G_{422} = 0$
(the corresponding complex conjugate terms are
$\G_{414} = 0$ and $\G_{411} = 0$).
The null tetrad $e_a^\m$ can be completed as follows:
\begin{eqnarray}
e^1 &=& d \z - Y d v; \\ \nonumber
e^2 &=& d \Z - \Y d v; \\ \nonumber
e^4 &=&  d v - h e^3.
\end{eqnarray}
The inverse tetrad has the form
\begin{eqnarray}
 \d_1 &=& \d_\z - \Y \d_u ; \\ \nonumber
\d_2 &=&  \d_\Z - Y \d_u ; \\  \nonumber
\d_3 &=&  \d_u - h \d_4 ;\\   \nonumber
\d_4 &=&  \d_v + Y \d_\z + \Y \d_\Z - Y  \Y \d_u .
\end{eqnarray}
It was shown in [8] that
$$
\G_{42} = \G_{42a} e^a  = - d Y - h Y,_4 e^4 .
$$
The congruence  $e^3 $ is geodesic if $ \G_{424} =
-Y,_4 (1-h) = 0, $
and is shear free if $ \G_{422} = -Y,_2 = 0.$
Thus  the function $ Y $ with the conditions
$$
Y,_2 = Y,_4 = 0,
$$
defines a shear free and geodesic congruence.
\vfill\eject

\newpage
\begin{thebibliography}{99}
\bibitem{[1]}
Cook, J.B. {\it et al}
(The Binary Black Hole Grand Challenge Alliance)
Phys. Rev. Lett. {\bf 80} (1998) p.2512.
\bibitem{[2]}
T. Dray and G. 't Hooft,
Nucl. Phys. B {\bf 253} (1985) 173
\bibitem{[3]}
M. Fabbichesi, R. Pettorino, G. Veneziano and G.A. Vilkovisky,
Nucl. Phys. B {\bf 419} (1994)147
\bibitem{[4]}
P.C.Aichelburg and R.U. Sexl, Gen. Rel.Grav. {\bf 2} (1971) 303
\bibitem{[5]}
V. Ferrari and P. Pendenza, Gen. Rel. Grav. {\bf 22} (1990) 1105
\bibitem{[6]}
C.O.Luosto and N. Sanchez, Nucl. Phys. B {\bf 355} (1991) 231
\bibitem{[7]}
C.O.Luosto and N. Sanchez, Nucl. Phys. B {\bf 383} (1992) 377
\bibitem{[8]}
G.C. Debney, R.P. Kerr and  A. Schild, J. Math. Phys. {\bf10} (1969) 1842.
\bibitem{[9]}
D.Kramer, H.Stephani, E. Herlt, M.MacCallum, {\it Exact Solutions
of Einstein's Field Equations,} Cambridge Univ. Press, Cambridge 1980.
\bibitem{[10]}
R. Penrose, J. Math. Phys. {\bf8} (1967) 345.
\bibitem{[11]}
D.Cox and E.J. Flaherty, Comm. Math. Phys. {\bf 47} (1976) 75.
\bibitem{[12]}
E.T. Newman,
J. Math. Phys. {\bf14} (1973) 102.
\bibitem{[13]}
R.W. Lind and E.T. Newman,
J. Math. Phys. {\bf 15} (1974) 1103.
\bibitem{[14]}
D. Ivanenko and A.Ya. Burinskii,
Izvestiya Vuzov Fiz. $N^0$ 7 (1978) 113 (Sov. Phys. J. USA)
\bibitem{[15]}
A. Burinskii, R.P. Kerr, Z. Perj\'es,
Preprint gr-qc/9501012
\bibitem{[16]}
E.Brinis Udeschini and G. Magli,
J. Math. Phys. {\bf 37} (1996), 5695
\bibitem{[17]}
R.P.Kerr and  W.B. Wilson,
Gen. Rel. Grav. {\bf 10} (1979) 273
\bibitem{[18]}
H. Balasin and H. Nachbagauer, Class. Quantum Grav. {\bf 12} (1995) 707
\bibitem{[19]}
H. Balasin and H. Nachbagauer, Class. Quantum Grav. {\bf 13} (1996) 731
\bibitem{[20]}
A.Sen,
Phys. Rev. Lett. {\bf 69} (1992) 1006
\bibitem{[21]}
A.Burinskii,
Phys. Rev. D {\bf 52} (1995) 5826
\bibitem{[22]}
A. Burinskii
and G. Magli,
in ``{\it Internal Structure of Black Holes
and Spacetime Singularities}", L. Burko and A. Ori eds.,
Ann. Israel Phys. Soc. {\bf 13} (IOP publ., Bristol).
\end{thebibliography}
\end{document}